\documentclass[journal]{IEEEtran}

\usepackage{amssymb}
\usepackage{color}

\ifCLASSINFOpdf
\else
   \usepackage[dvips]{graphicx}
\fi
\usepackage{url}

\usepackage{amsmath}
\usepackage{graphicx}
\usepackage{array}
\usepackage{xcolor}
\begin{document}

\title{GNCformer: Enhanced Self-attention for Automatic Speech Recognition}

\author{ Junhua Li, Zhikui Duan, Shiren Li, Xinmei Yu, Guangguang Yang
\thanks{This work has been submitted to the IEEE for possible publication. Copyright may be transferred without notice, after which this version may no longer be accessible.}

\thanks{Shiren Li is from Sun Yat-Sen University, Guangzhou, P.R.China. (e-mail: lishr6@mail3.sysu.edu.cn)}

\thanks{This work was supported by Research Foundation No.2020A1515111107, No.2021B1515120025 and No.2021KSYS008.}}

\markboth{IEEE Signal Processing Letters}
{Shell \MakeLowercase{\textit{et al.}}: Bare Demo of IEEEtran.cls for IEEE Journals}
\maketitle

\begin{abstract}
In this paper, an Enhanced Self-Attention (ESA) mechanism has been put forward for robust feature extraction. The proposed ESA is integrated with the recursive gated convolution and self-attention mechanism. In particular, the former is used to capture multi-order feature interaction and the latter is for global feature extraction. In addition, the location of interest that is suitable for inserting the ESA is also worth being explored. In this paper, the ESA is embedded into the encoder layer of the Transformer network for automatic speech recognition (ASR) tasks, and this newly proposed model is named GNCformer. The effectiveness of the GNCformer has been validated using two datasets, that are Aishell-1 and HKUST. Experimental results show that, compared with the Transformer network, 0.8\%CER, and 1.2\%CER improvement for these two mentioned datasets, respectively, can be achieved. It is worth mentioning that only 1.4M additional parameters have been involved in our proposed GNCformer.
 
\end{abstract}

\begin{IEEEkeywords}
speech recognition; Transformer; 
enhance self-attention; multi-order interaction
\end{IEEEkeywords}

\IEEEpeerreviewmaketitle

\section{Introduction}
Transformer \cite{speechtransformer}, as an attention-based encoder-decoder network, has been widely used in various application scenarios, such as MultiModal Machine Learning (MMML) \cite{Vilt} \cite{ViLBERT}, Natural Language Processing (NLP) \cite{transformer_XL} \cite{he2022language} and Computer Vision (CV) \cite{zhao2022hybrid} \cite{liu2021swin}. Similar with other fields, Transformer has also been introduced to ASR \cite{gulati2020conformer} \cite{peng2022branchformer} \cite{LFEformer} community for its excellent model performance and has become one of the most popular approaches in this field. Current research attention is focused on performance enhancement, like the local feature enhancement \cite{ETC}. Such kind of approaches contributes a positive effect to the development of the ASR.

Recently, the LSRA module  proposed in \cite{Lite} adopts a dual-branch structure in which one branch utilizes an attention mechanism to extract global information, while the local information is extracted by the other branch under the help of convolution. These pieces of information are then fed into the Feed-Forward Network (FFN) layer for fusion in the Transformer network. The LFEformer model proposed in \cite{LFEformer}  can further improve the model performance with a similar dual-branch scheme by extracting robust local features with a sliding window of varying sizes. With the parallel-branch scheme, Branchformer \cite{peng2022branchformer} employs a cgMLP module for local feature extraction to further improve the model performance.
\begin{figure}[!t]
   \centering
    \includegraphics[width=0.49\textwidth]{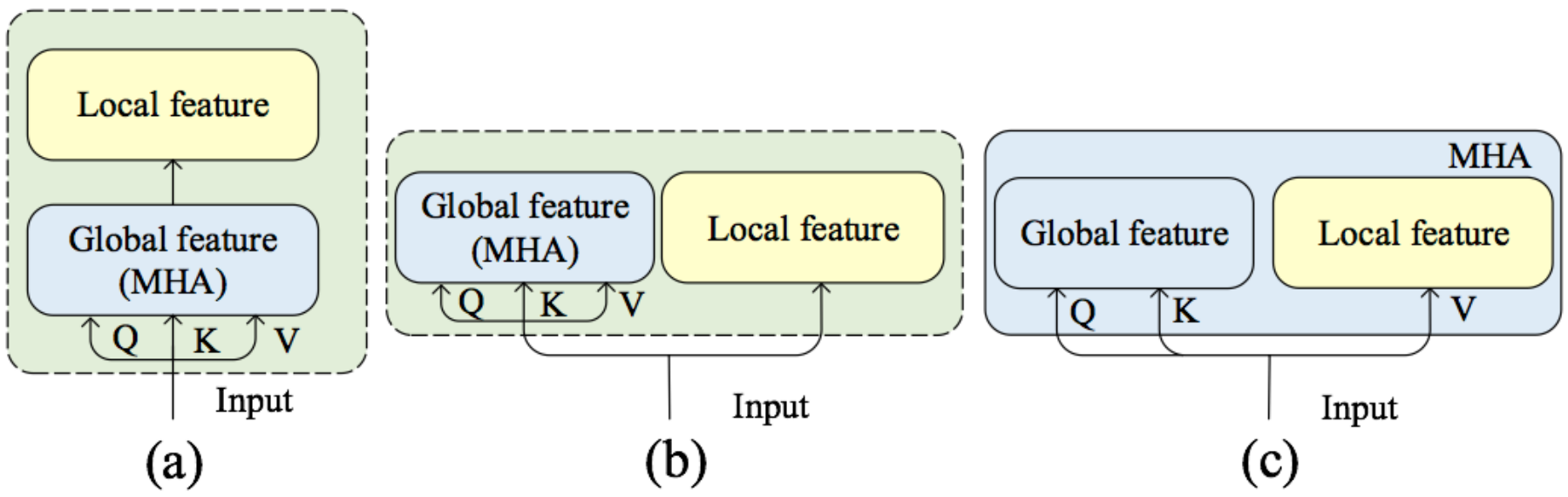}
    \vspace{-20pt}
    \caption{Integration of local features and global features. (a) represents the serial fusion of global and local features; (b) represents the parallel fusion of global and local features; (c) is the novel fusion method proposed for integrating global and local features in this paper. In (a) and (b), the output of MHA is viewed as the global features, so the integration operation between the local and global features is conducted outside the MHA. However, the multiplication between $Q$ and $K$, regarded as the attention matrix, is viewed as the global feature in the proposed model, so the integration operation is conducted inside the MHA.}
    \label{3di}
   \vspace{-10pt}
\end{figure}
However, these attempts in open literature tend to enhance local features in Transformer-based models but fail to pay attention to the multi-order feature interactions, which is of utmost importance for model improvement. 

To address the above-mentioned problem, an ESA mechanism has been proposed in this study. In this approach, a recursive gated convolution (abbreviated as $g^nConv$) \cite{HorNet} that utilizes multi-order interaction among local features has been integrated with a self-attention mechanism for the purpose of robust feature extraction. 
In particular, the $g^nConv$ is embedded after the linear projection of the input, which is $V$ in the self-attention mechanism. 
This novel approach allows the $g^nConv$ mechanism to be smoothly fused with the self-attention mechanism so that the ESA can be successfully constructed. Moreover, it is worth mentioning that this structure integrates local features with global ones inside multi-head self-attention (MHA). This is different from the existing approaches since the integration of these two kinds of features is always outside the MHA, shown in Fig.\ref{3di}.

Another concern in this study, as a critical factor in determining model performance, is how to select the suitable position for inserting ESA. In this paper, the encoder layer in Transformer is selected as the position of interest.

\begin{figure*}[!t]
   \centering
  \begin{minipage}[t]{0.45\linewidth}
    \includegraphics[width=\textwidth]{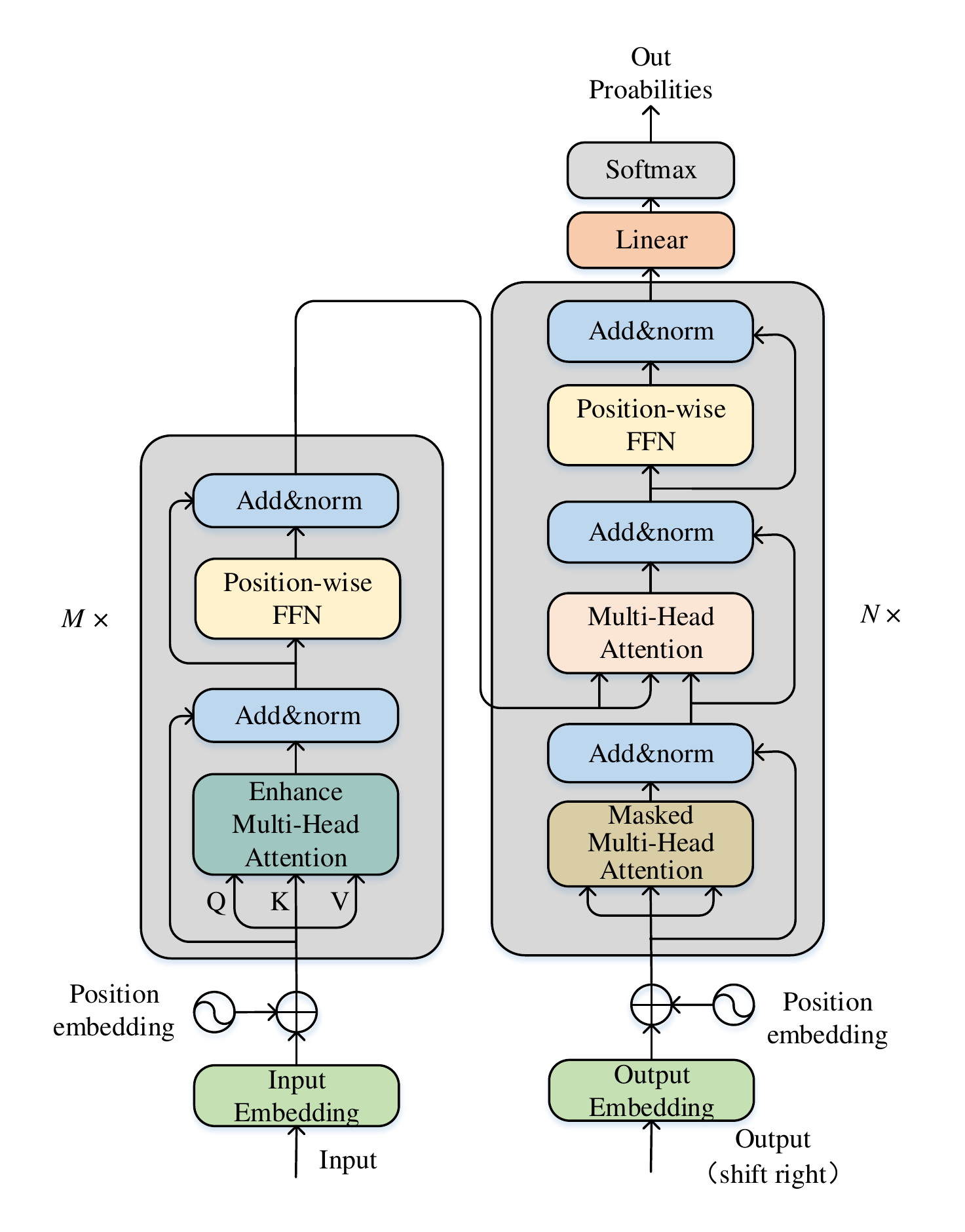}
    \centerline{(a)}
    \end{minipage}
    \begin{minipage}[t]{0.53\linewidth}
 \includegraphics[width=\textwidth]{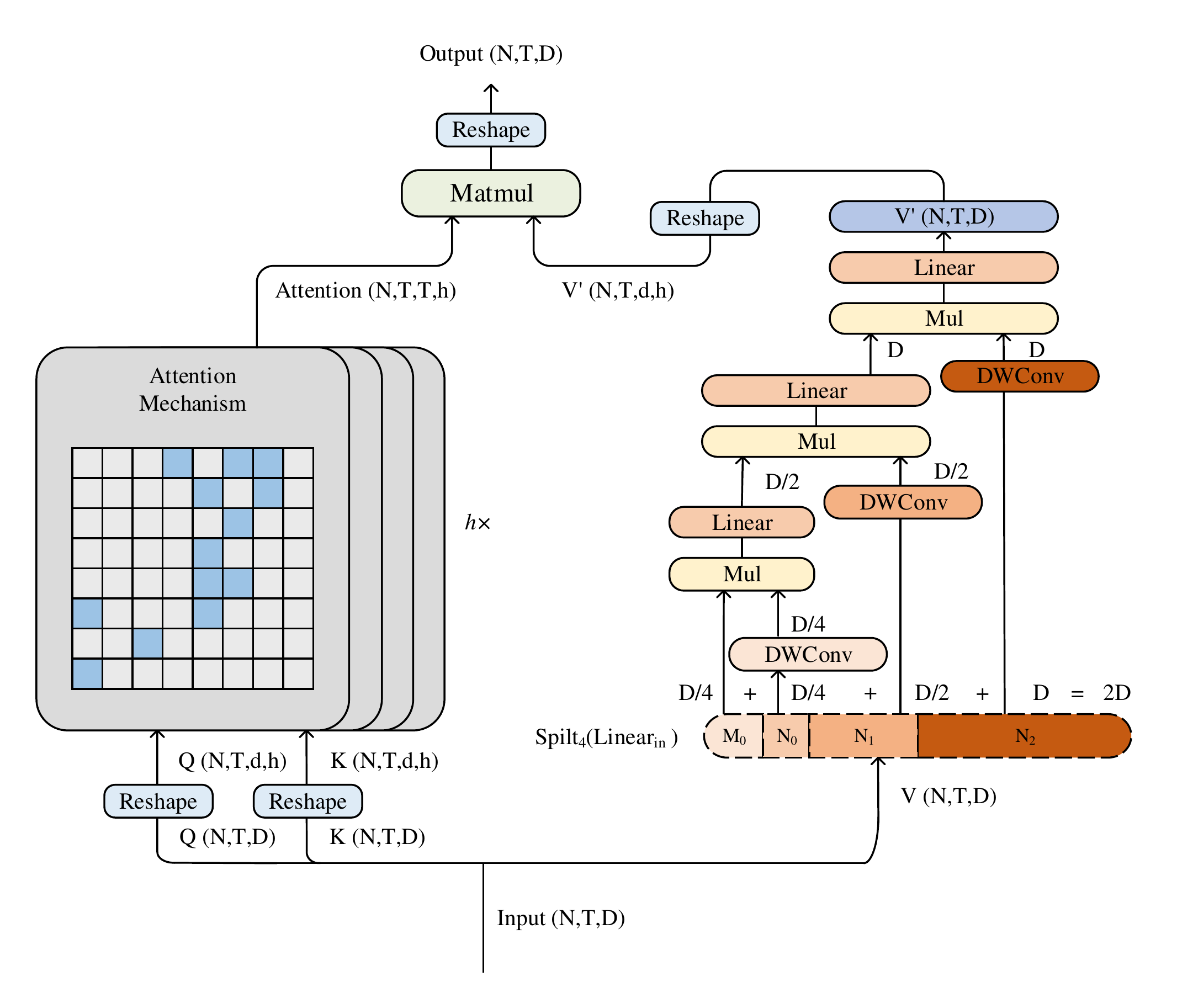}
    \centerline{(b)}
   \end{minipage}
\caption{The structure of the proposed network. In this figure, the $order$ is set to 3 for simplicity.  (a) is the overall architecture of the proposed GNCformer network. The GNCformer consists of two parts, that are, the encoder and decoder. (b) is the structure of the enhanced self-attention mechanism. `DWConv' represents depth-wise separable convolution. $D$ is the dimension of the input tensor and $h$ is the number of heads. $D=h\times d$. `Mul' represents element-wise multiplication between tensors. 
`Matmul' represents matrix multiplication. Matrix multiplication of Query ($Q$) and Key ($K$) tensors generates an attention matrix. The tensor $V^{'}$, obtained by using a recursive gated convolution mechanism, is the multi-order local feature interaction. Finally, the output of the enhanced self-attention mechanism is obtained by multiplying the attention  matrix and the $V^{'}$ tensor.}
    \label{overall}
    \vspace{-10pt}
\end{figure*}

In short, a network called GNCformer has been proposed for ASR tasks, in which the ESA is inserted into the encoder layer of the Transformer model. The proposed ESA is the integration between self-attention and $g^nConv$. The proposed GNCformer is tested on two widely-used datasets, which are Aishell-1 and HKUST, respectively. Experimental results show that the performance of the GNCformer outperforms other popular models on both datasets. Their character error rate (CER) is $0.6\%/0.7\%$ and $1.0\%$ better than the Transformer baseline when the language module (LM) is excluded and $0.6\%/0.8\%$ and $1.2\%$ when LM is included.
The main contributions of this study are listed as follows:
\begin{itemize}
    \item An enhanced self-attention mechanism has been proposed to extract the robust features. It contains the recursive gated convolution and self-attention. The former is used to extract the multi-order interaction among local features and the latter is used for global feature extraction.
    \item The proposed ESA is inserted into the encoder layer of the Transformer for ASR tasks and this integrated network is called GNCformer.   
    \item The effectiveness of the proposed GNCformer has been validated using two widely-used datasets, that are Aishell-1 and HKUST. Compared with the baseline Transformer model and other popular models, the GNCformer model has better performance.
\end{itemize}
\section{Methodology}
 
In this section, the overall structure of the GNCformer network is introduced first. Then the details of the major component, which is the ESA, are presented. 

 \subsection{Overall Structure of GNCformer}\label{sec3_1}
As shown in Fig. \ref{overall} (a), the proposed GNCformer consists of two components, which are the encoder and decoder. 
Each encoder layer consists of two modules, which are a proposed ESA module and a FFN module. Each module also incorporates residual connections and layer normalization within the model structure. The ESA module is shown in Fig. \ref{overall} (b), which includes a self-attention mechanism and a recursive gated convolution.
In particular, the ESA model fuses the local feature interaction extracted by $g^nConv$ with the global features provided by the self-attention mechanism. The integration of these two kinds of features allows better feature extraction to be achieved than the existing methods.
The input $X$ is linearly mapped to the matrices $Q$, $K$, and $V$. The attention matrix is then computed by performing a matrix multiplication of $Q$ and $K$. Then $Softmax(.)$ function is conducted on the attention matrix. The attention matrix captures long-range interactions among tokens. $V$ is used as the input of recursive gated convolution to capture the multi-order local feature interaction. Thus, the expression for fusing multi-order local feature interaction with the self-attention mechanism can be defined as:
\begin{equation}\label{eq_Dk}
  ESA = Attention * g^nConv(V),
\end{equation} 
\begin{equation}\label{eq_attn}
    Attention=Softmax[\frac{QK^T}{\sqrt{d_{k}}}],
\end{equation}
where $g^nConv(.)$ is the representation of recursive gated convolution. 
The decoder layer is similar with the Transformer baseline, and the details can refer to \cite{speechtransformer}.

\subsection{Enhanced Self-attention Mechanism}\label{sec3_2}
ESA is a method that smoothly incorporates the $g^nConv(.)$ mechanism into the self-attention mechanism. In the following, we will explain the detailed effects of ESA from the perspectives of the $g^nConv(.)$ mechanism.

The basis of the $g^nconv$ is the gated convolution ($gConv$). We first introduce the details of the $gConv$ and then generalize it to the $g^nConv$.  Let $\textbf{x}=(x_1,…,x_T)$ be the input features, so the output features $\textbf{y}=gConv(\textbf{x})$ can be represented as follows:
\begin{equation}\label{eq_decoder_out}
   [M_0^{(T,D)},N_0^{(T,D)}] = Spilt_2[Linear_{in}(\textbf{x})] \in 
   \mathbb{R}^{(T,2D)},
\end{equation}
\begin{equation}\label{eq_decoder_out}
   M_1^{(T,D)} = DWConv(N_0) \odot  M_0 
   \in 
   \mathbb{R}^{(T,D)}, 
\end{equation}
\begin{equation}\label{eq_decoder_out} \textbf{y}= Linear(M_1) 
   \in 
   \mathbb{R}^{(T,D)}.
\end{equation}

The dimension of input features is increased from 1D to 2D by $Linear_{in}$, resulting in 2D output features.
The output of $Linear_{in}$ is separated into $M_0^{(T,D)}$ and $N_0^{(T,D)}$ using the $Split_2$ function. 
$DWConv$ is a depth-wise convolution used for local information extraction here. 
In addition, the adjacent features can achieve information interaction via element-wise multiplication. Therefore, $gConv$ is suitable for local feature interaction in which each $M_0^{(i)}$ has interacted with its neighbor feature $N_0^{(j)}$. 
$g^nConv()$ is a recursive extension of the $gConv$ function, performing convolution through a recursive design. In mathematics, this means that the input consists of $M_0$ and ${N_{k=0}^{n-1}}$.
\begin{equation}\label{eq_linear_in}
    [M_0^{(T,D_0)},N_0^{(T,D_0)},...,N_{n-1}^{(T,D_{n-1})}] = Spilt_{n+1}[Linear_{in}(V)], 
\end{equation}   
\begin{equation}\label{eq_gnconv}
  Linear_{in}(V)  \in \mathbb{R}^{(T,D_0 + \sum_{0 \le k \le n-1}D_k)},
\end{equation} 
\begin{equation}\label{eq_gnconv}
\begin{split}
      M_{k+1} = DWConv_k(N_k) \odot Linear_{D_{k-1}\rightarrow D_k}(M_k) /\alpha,  \\ 
      k=0,...,n-1, 
\end{split}
\end{equation}
where $\frac{1}{\alpha}$ is a scaling factor that stabilizes the training. $DWConv_k$ are a set of depth-wise convolution layers in different orders.
The function $Split_n$ divides the output of $Linear_{in}(V)$ into $n+1$ parts, each of which has different dimensional sizes.
From the recursive formula Equation \ref{eq_gnconv}, it is apparent that the interaction order of $M_k$ will be increased by 1 after each step. 
$Linear_{D_{k-1}\rightarrow D_k}(M_k)$ is used to match the dimension in different orders. To guarantee there is not too much computation increase involved in the high-order interactions, the dimension $D_k$ in each order can be written as:
\begin{equation}\label{eq_Dk}
  D_k = \frac{D}{2^{n-k-1}}, \quad 0 \le k \le n-1.
\end{equation} 

Finally, the two different kinds of features are then fused together using matrix multiplication. 

\section{Experiment}

\subsection{Experimental Setting}\label{sec4_1}
In this paper, the proposed GNCformer network is tested in the ASR tasks using two widely-used datasets, that are Aishell-1 \cite{Aishell1} and HKUST \cite{liu2006hkust}. All the experiments are implemented on the NVIDIA Tesla T4 GPU with ESPNet toolkit \cite{watanabe2018espnet}. The number of encoder layers and decoder layers is set to $M$=$N$=6. In the Recursive Gated Convolution part, the input dimension is 256 with the order of 5. After the first linear mapping, the input dimension is extended to 512. Other parameters follow the settings given in \cite{LFEformer}.

\subsection{Comparison with Other Popular Modles}\label{sec4_2}
The comparison among the proposed GNCformer network and other popular models is shown in TABLE \ref{table_aishell} and TABLE \ref{table_hkust}. Obviously, in the Aishell-1 dataset, GNCformer without LM has 0.6\% and 0.7\% CER improvement in the development set and testing set, respectively, in comparison with the Transformer baseline. 0.6\% CER improvement on Dev and 0.8\% CER development on Testing can be achieved when LM is involved. In the HKUST dataset, the GNCformer network is 1.0\% CER and 1.2\% CER better than the Transformer network when LM is correspondingly excluded and included. Model size of baseline and the proposed GNCformer is listed in both tables. From these two tables, one can note that the GNCformer can achieve impressive results with just an additional 1.4M parameters. Moreover, the GNCformer surpasses other popular models listed in TABLE \ref{table_aishell} and TABLE \ref{table_hkust}.

\begin{table}[!htbp]
	\centering
\caption{Comparison among GNCformer and other ASR models in Aishell-1 dataset} 
		\vspace{2pt}
		\setlength{\tabcolsep}{3mm}{
	\begin{tabular}{lccc}
		\hline
    		Model&Dev (\%)  &Test (\%) &Size(M)\\
		
		\hline
		Transformer without LM \cite{transformer2017}       & 5.8        & 6.3 &22.47      \\
		Transformer with LM \cite{transformer2017}       & 5.6       & 6.0  &22.47     \\
		\hline
        Chunk-Flow SA-T \cite{Transducers}  & 8.58        & 9.80   & -   \\
        Sync-Transformer \cite{Tian2020SynchronousTF}   & 7.91        &  8.91    &- \\
	Masked-NAT \cite{chen2020non}  &      6.4&   7.1 &-\\
	Insertion-NAT \cite{2353803bf0d24fddaf55abd105215289} & 6.1  &  6.7 &-\\
  	ESPNet-RNN \cite{20dsad19} & 6.8 & 8.0 &-\\	
        AT \cite{9413429}&   5.5  &   5.9 &-\\
        Realformer with LM \cite{he2020realformer} & 5.7 &6.1 &-\\
        LASO \cite{bai2020listen}          &5.8 &  6.4&-\\
        DRTNet with LM \cite{DRTNet}       &5.3 &5.6&-\\
    	\hline
    	GNCformer without LM                     & \textbf{5.2} & \textbf{5.6} &23.93    \\
    	GNCformer with LM                               & \textbf{5.0} & \textbf{5.2} &23.93    \\
		\hline\\
	\end{tabular}}
	\label{table_aishell}
	\vspace{-15pt}
\end{table} 

\begin{table}[tp]
	\centering
		\caption{Performance of ASR models in HKUST dataset.} 
		\vspace{-2pt}
	\begin{tabular}{lccc}
		\hline
		Model & CER (\%) & size(M)\\		
		\hline
        CTC with LM \cite{7472152} & 34.8& -\\
        Self-attention Aligner \cite{Dong2019SelfattentionAA} & 24.1 &-\\
        Chain-TDNN \cite{povey16_interspeech} & 23.7 &-\\
        Extended-RNA \cite{dong2019extending} & 26.6 &-\\
        Joint CTC-attention model/ESPNet \cite{7953075} & 27.4 &-\\
		SAM with LM \cite{Dong2019SelfattentionAA}& 24.92&-\\
        Transformer with LM \cite{transformer2017} & 22.6&21.95\\
        Transformer without LM \cite{transformer2017} & 22.6&21.95\\
        DRTNet without LM \cite{DRTNet} &22.4&-\\
        \hline
        GNCformer with LM & \textbf{21.4} &23.43\\
        GNCformer without LM & \textbf{21.6} &23.43\\
        \hline\\
	\end{tabular}
	\label{table_hkust}
	\vspace{-10pt}
\end{table} 

\begin{table}[tp]
    \centering
        \caption{results in different orders of feature interaction. $\bigtriangledown$ means the difference of CER between the proposed GNCformer with the given order and the Transformer model.}
    \begin{tabular}{lccccc}
       \hline
       $order$  &3 &5 &7 &9\\
       \hline
        Num Params(M) &23.92 &23.93 &23.94 &23.94\\
        Input dims(D) &256    &256    &256    &256\\

        Attetnion Heads &4  &4 &4 &4 \\
        Conv Kernel Size &32 &32 &32 &32 \\
        
        \hline
        CER(Aishell,\%)&5.5/5.8  &\textbf{5.2/5.6} &5.3/5.7&5.3/5.7  \\
        $\bigtriangledown$ (Aishell)  &0.3/0.5  &0.6/0.7 &0.5/0.6 &0.5/0.6 \\
        \hline 
        CER(HKUST,\%) &\textbf{21.3}\ &21.6 &21.8  &21.8 \\
        $\bigtriangledown$  (HKUST)  &1.3 &1.0 &0.8 &0.8 \\
        \hline 
    \end{tabular}
    \label{differenet_order}
\end{table}

\begin{table}[tp]
    \centering
        \caption{The separation of the input of recursive convolution at different orders.}
    \begin{tabular}{l|cccccccccc}
       \hline
       order & \multicolumn{9}{c}{[$N_{n-1},...,N_1,N_0,M_0$]}  \\
       \hline
       3  &\makebox[0.02\textwidth][c]{[256,}  &\makebox[0.01\textwidth][c]{128,}&\makebox[0.01\textwidth][c]{64,} &\makebox[0.01\textwidth][c]{64]} & & & & &&\\
       5 &\makebox[0.02\textwidth][c]{[256,} &\makebox[0.01\textwidth][c]{128,}&\makebox[0.01\textwidth][c]{64,}&\makebox[0.01\textwidth][c]{32,}&\makebox[0.01\textwidth][c]{16,}&\makebox[0.01\textwidth][c]{16]} & & & &\\
       7  &\makebox[0.02\textwidth][c]{[256,} &\makebox[0.01\textwidth][c]{128,}&\makebox[0.01\textwidth][c]{64,}&\makebox[0.01\textwidth][c]{32,}&\makebox[0.01\textwidth][c]{16,}&\makebox[0.01\textwidth][c]{8},&\makebox[0.01\textwidth][c]{4},&\makebox[0.01\textwidth][c]{4]} & &\\
       9 &\makebox[0.01\textwidth][c]{[256,} &\makebox[0.01\textwidth][c]{128,}&\makebox[0.01\textwidth][c]{64,}&\makebox[0.01\textwidth][c]{32,}&\makebox[0.01\textwidth][c]{16,}&\makebox[0.01\textwidth][c]{8,}&\makebox[0.01\textwidth][c]{4,}&\makebox[0.01\textwidth][c]{2,}&\makebox[0.01\textwidth][c]{1}, &\makebox[0.01\textwidth][c]{1]}\\ 
        \hline 
    \end{tabular}
    \label{differenet_dims}
\end{table}

\subsection{Ablation Studies}\label{sec4_4}

The specific contribution of the ESA is demonstrated in this part. First, how the order of interaction affects the model performance is analyzed. Then we focus on exploring where is the optimal location and which is the best way for ESA embedding. It should be mentioned that the GNCformer implemented in this part is without LM. In $g^nConv$, the dimension of output features is twice that of input features for the tradeoff between computation efficiency and model performance.

First, exploration is focused on how the order of ESA actually affects the GNCformer model performance. The corresponding experiments are conducted on both the Aishell-1 dataset and the HKUST dataset. From Table \ref{differenet_order}, one can note that the model performance varies with different orders and all of them are better than the Transformer baseline. 
The ESA achieved its best results on the Aishell-1 dataset at $order$=5, leading to a 0.6\%/0.7\% CER improvement over the baseline. With regards to the HKUST dataset, the best performance occurred at $order$=3 with 1.3\% CER improvement over the baseline. In order to unify the settings in both datasets, $order=5$ is selected in this study. 
Moreover, the input of recursive convolution at different orders has been separated and their dimensional sizes are listed in Table \ref{differenet_dims}. When the order is set to 7 or 9, features in the first iteration, which are $M_0$ and $N_0$, contain fewer dimensions (such as 4 in order 7 and 1 in order 9) compared with smaller orders. Thus, the subsequent interaction between features with less dimension may lead to sub-optimal interactions.

In addition, the location of interest for embedding the proposed ESA has been investigated. The candidate locations are the encoder layer, the decoder layer, and both the encoder layer and decoder layer in the Transformer network. The corresponding experiments are conducted on the Aishell-1 dataset and the experimental results are shown in TABLE \ref{ESDSinHKUST}. When ESA is embedded into the encoder and decoder layer, we call them ESA\_EN and ESA\_DE for short. 
Based on the experimental results, one can note that all the results for the model with embedded ESA have better performance than that without ESA. When the proposed ESA is used separately in the encoder layer and decoder layer, they can bring improvement to the model by 0.6\%/0.7\% CER and 0.2\%/0.2\% CER, respectively. Compared with the baseline, the model exhibits 0.4\%/0.4\% improvement when the ESA is jointly used with both the encoder layer and decoder layer. Thus, the encoder layer, rather than the decoder or both the encoder layer decoder layer, is selected for ESA embedding in this study.

Finally, we have also looked into how to embed the ESA appropriately. Three structures are chosen for testing, as shown in Fig.\ref{3di}. In Fig.\ref{3di} (a), the output of MHA is fed into the $g^nConv$ module. In Fig.\ref{3di} (b), the input is fed into the MHA and $g^nConv$ module simultaneously and the sum of their results is regarded as the final output. It should be noted that these two kinds of structures view the MHA as a whole and integrate the MHA and $g^nConv$ with parallel or serial connection. Different from (a) and (b), the $g^nConv$ is integrated with self-attention in MHA in (c). All the experimental results are shown in TABLE \ref{different}. It is apparent that all of them have better model performance than the Transformer baseline, indicating that the proposed ESA is useful for model improvement. Moreover, the comparison shows that our structure provides the best results. It has 0.4\%/0.4\% CER better than the parallel connection and 0.3\%/0.3\% CER than the serial connection. These results demonstrate that the structure with embedded ESA in the MHA is better than those that regard HMA as a whole.

\begin{table}[!htbp]
    \centering
    \caption{The application of the Enhance Attention module in the encoder and decoder attention modules.}
       \begin{tabular}{lc|ccc}
       \hline
       ESA\_EN &ESA\_DE &CER(Aishell-1) \\
        \hline
        &  &5.8\%/6.3\%\\
       \hline
      \checkmark & &5.2\%/5.6\%\\
      \hline
      & \checkmark &5.6\%/6.1\%\\
      \hline
      \checkmark &\checkmark & 5.4\%/5.9\% \\
        \hline 
    \end{tabular}
    \label{ESDSinHKUST}  
\end{table}

\begin{table}[!htbp]
    \centering
    \caption{Experimental results of GNCformer using different fusion methods in Aishell-1 data sets}
       \begin{tabular}{l|ccc}
       \hline
       Meathod &Parallel &Serial &Our\\
       \hline
       CER(\%) &5.6/6.0 &5.5/5.9 &5.2/5.6\\
     
        \hline 
    \end{tabular}
    \label{different}  
\end{table}

\section{Conclusion}
The ESA mechanism proposed in this paper integrates recursive gated convolution and self-attention for better feature extraction compared with existing models. Such ESA can not only acquire global information interaction captured by self-attention mechanism but also extract multi-order local information interaction captured by recursive gated convolution. Thus, the GNCformer model, one of whose core components is ESA, is able to capture robust features from the input data. In addition, the validity of the GNCformer model has been comprehensively verified on two widely used Aishell-1 and HKUST datasets. In particular, the GNCformer can improve model performance with negligible additional model parameters.

\end{document}